
\overfullrule=0pt
\magnification=\magstep1
\looseness=2
\hsize=6.5 truein
\vsize=9.0 truein
\baselineskip=14pt plus2pt
\parskip=6pt plus4pt minus2pt

\font\bigrm=cmr10 scaled \magstep2
\newcount\qno
\newcount\eqA
\newcount\eqB
\newcount\eqC
\newcount\eqD
\newcount\eqE
\newcount\eqF
\def\Eqn{(\the\qno)}
\def\nxt{\global\advance\qno by1 \eqno\Eqn}
\def\EQA{\advance \eqA by -\eqA \advance \eqA by \qno}
\def\EQB{\advance \eqB by -\eqB \advance \eqB by \qno}
\def\EQC{\advance \eqC by -\eqC \advance \eqC by \qno}
\def\EQD{\advance \eqD by -\eqD \advance \eqD by \qno}
\def\EQE{\advance \eqE by -\eqE \advance \eqE by \qno}
\def\EQF{\advance \eqF by -\eqF \advance \eqF by \qno}
\newcount\plt
\newcount\plA
\newcount\plB
\newcount\plC
\newcount\plD
\newcount\plE
\newcount\plF
\newcount\plG
\newcount\plH
\newcount\plI
\newcount\plJ
\newcount\plK
\newcount\plL
\newcount\plM
\newcount\plN
\newcount\plO
\newcount\plP
\def\Plot{\the\plt\thinspace}
\def\nxtplt{\global\advance\plt by1 \Plot}
\def\PLOTA{\advance \plA by -\plA \advance \plA by \plt}
\def\PLOTB{\advance \plB by -\plB \advance \plB by \plt}
\def\PLOTC{\advance \plC by -\plC \advance \plC by \plt}
\def\PLOTD{\advance \plD by -\plD \advance \plD by \plt}
\def\PLOTE{\advance \plE by -\plE \advance \plE by \plt}
\def\PLOTF{\advance \plF by -\plF \advance \plF by \plt}
\def\PLOTG{\advance \plG by -\plG \advance \plG by \plt}
\def\PLOTH{\advance \plH by -\plH \advance \plH by \plt}
\def\PLOTI{\advance \plI by -\plI \advance \plI by \plt}
\def\PLOTJ{\advance \plJ by -\plJ \advance \plJ by \plt}
\def\PLOTK{\advance \plK by -\plK \advance \plK by \plt}
\def\PLOTL{\advance \plL by -\plL \advance \plL by \plt}
\def\PLOTM{\advance \plM by -\plM \advance \plM by \plt}
\def\PLOTN{\advance \plN by -\plN \advance \plN by \plt}
\def\PLOTO{\advance \plO by -\plO \advance \plO by \plt}
\def\PLOTP{\advance \plO by -\plO \advance \plO by \plt}
  \def\ps{\noindent\goodbreak
      \parshape 2 0truecm 15truecm 2truecm 13truecm}
  \def\ref#1;#2;#3;#4;#5.{\ps #1\ #2, {#3},\ {#4}, #5}
  \def\\{\hfil\break}
  \def\aa{Astron. \& Astrophys.}
  \def\apj{Ap. J.}
  \def\apjl{Ap. J. Lett.}
  \def\apjs{Ap. J. Suppl.}
  \def\mnras{M.N.R.A.S.}
  \def\plb{Phys. Lett. B}
  \def\prd{Phys. Rev. D}
  \def\prl{Phys. Rev. Lett.}
  
\def\Msun{~M_{\odot}\ }
\def\Mpc{~{\rm Mpc}\ }

\def\etal{{\it et al.}\ }
\def\kms{~\hbox{km\ s}^{-1}}
\def\M10{{\times 10^{10} M_{\odot}\ }}

\def\rms{{\rm rms}}
\def\kms{\ {\rm km}\ {\rm s}^{-1}}

\centerline{\bigrm Structure Formation with}
\centerline{\bigrm Cold + Hot Dark Matter}

\bigskip
\centerline{Anatoly Klypin\footnote*{Present address:
Department of Astronomy, New Mexico State University,
Box 30001/Dept. 4500, Las Cruces, NM 88003}}
\centerline{Department of Physics and Astronomy, University of Kansas,
Lawrence, KS 66045}

\medskip
\centerline{Jon Holtzman}
\centerline{Lowell Observatory, Mars Hill Road, Flagstaff,
AZ 86100}

\medskip
\centerline{Joel Primack and Enik\H{o}
Reg\H{o}s\footnote\dag{Present address: Institute of Astronomy,
Madingley Road, Cambridge CB3 0HA ENGLAND} }
\centerline{Physics Department, University of California,
Santa Cruz, CA 95064}

\bigskip
\centerline{\bf Abstract}

    We report results from high-resolution particle-mesh
(PM) N-body simulations of structure formation in an
$\Omega=1$ cosmological model with a mixture of Cold plus
Hot Dark Matter (C+HDM) having $\Omega_{\rm cold}=0.6$,
$\Omega_\nu=0.3$, and $\Omega_{\rm baryon}=0.1$. We present
analytic fits to the C+HDM power spectra for both
cold and hot ($\nu$) components, which provide initial
conditions for our nonlinear simulations. In order to
sample the neutrino velocities adequately, these simulations
included six times as many neutrino particles as cold
particles.  Our simulation boxes were 14, 50, and 200~Mpc
cubes (with $H_0=50$ km s$^{-1}$ Mpc$^{-1}$); we also did
comparison simulations for Cold Dark Matter (CDM) in a
50~Mpc box.

    C+HDM with linear bias factor $b=1.5$ is consistent both
with the COBE data and with the galaxy correlations we
calculate.  We find the number of halos as a function of
mass and redshift in our simulations; our results for both
CDM and C+HDM are well fit by a Press-Schechter model.  The
number density of galaxy-mass halos is smaller than for CDM,
especially at redshift $z>2$, but the numbers of
cluster-mass halos are comparable.  We also find that on
galaxy scales the neutrino velocities and flatter power
spectrum in C+HDM result in galaxy pairwise velocities that
are in good agreement with the data, and about 30\% smaller
than in CDM with the same biasing factor.  On scales of
several tens of Mpc, the C+HDM streaming velocities are
considerably larger than CDM.  As a result, the ``cosmic Mach
number'' in C+HDM is about a factor of two larger than in
CDM, and probably in better agreement with observations.

    Thus C+HDM looks promising as a model of structure
formation.  The presence of a hot component requires the
introduction of a {\it single} additional parameter beyond
standard CDM --- the light neutrino mass, or equivalently
$\Omega_\nu$ --- and allows this model to fit essentially
{\it all} the available cosmological data remarkably well.
The tau neutrino is predicted to have mass of about 7 eV,
compatible with the MSW explanation of the solar neutrino
data together with a long-popular particle physics model.
We outline a number of additional tests to which the C+HDM
model should be subjected.

\bigskip

\noindent {\it Subject headings:} cosmology: theory ---
dark matter --- large scale structure of the universe ---
galaxies: formation --- galaxies: clustering

\bigskip

\centerline{To be published in The Astrophysical Journal, October 1993.}

\vfill\eject

\centerline{\bf 1. Introduction}

The cold dark matter (CDM) model is perhaps the simplest
potentially viable model for dark matter and structure
formation in the universe.  As is well known, CDM is based
on the following set of assumptions (Blumenthal \etal 1984):
the dark matter is cold, and the initial fluctuations are
adiabatic, Gaussian, and have a Zel'dovich spectrum, as
predicted by the simplest models of inflation. The
``standard'' CDM model makes the additional assumption that
$\Omega=1$, with $\Omega_{\rm baryon} \approx 0.05$ from
standard big bang nucleosynthesis (Walker \etal 1991) and
$\Omega_{\rm cold}=\Omega-\Omega_{\rm baryon}$; fitting
observational data then requires that the Hubble parameter
$h\equiv H_0/100\ \kms \Mpc\!^{-1} \approx0.5$ and that
galaxy formation be ``biased'' (Davis \etal 1985).  This
attractive model had great success, but it is now well known
to be in varying degrees of difficulty with many sets of
observational data (see e.g. Davis \etal 1992 and references
therein).

Perhaps the simplest variant of CDM that remains viable has
$\Omega\approx0.2$ with $h \approx 1$ and a cosmological
constant $\lambda \equiv \Lambda/3H_0^2 = 1-\Omega$ for
consistency with inflation and with CMB constraints.  This
model is claimed to be consistent with the APM galaxy
angular correlation function $w_g(\theta)$ (Efstathiou,
Sutherland, \& Maddox 1990), with the observed rich cluster
correlation function $\xi_c(r)$ (Holtzman \& Primack 1993),
mass function (Lilje 1992, Bacall \& Cen 1992), and power
spectra from clusters (Scaramella 1992), the CfA slices
(Vogeley \etal 1992), and the Southern Sky redshift survey
(Park, Gott, \& da Costa 1992).  However, there are several
indications that $\Omega \approx 1$, for example CMB dipole
vs. QDOT/IRAS data (Rowan-Robinson \etal 1990, Strauss \etal
1992), comparison of IRAS density and galaxy peculiar
velocity data (Kaiser \etal 1991, Dekel \etal 1992),
reconstructing Gaussian initial conditions from the POTENT
analysis of galaxy peculiar velocity data (Nusser and Dekel
1992), and void outflow (Dekel \& Rees 1992). While this
evidence that $\Omega=1$ is still not compelling, and the
arguments for a large Hubble parameter and an old universe
do point toward smaller $\Omega$, we will adopt $\Omega=1$
and $h\approx0.5$ as hypotheses in this paper.

The question arises whether {\it any} $\Omega=1$ model with
a physically motivated smooth spectrum of adiabatic Gaussian
fluctuations can account for all the data now available,
including the COBE CMB fluctuations (corresponding to
scales of 3000--300 h$^{-1}$ Mpc), large scale structure data
(300--10 h$^{-1}$ Mpc scales: galaxy angular correlations
$w_g(\theta)$, the cluster correlation function $\xi_c$, and
galaxy streaming velocities), and smaller scale structure
data (10 h$^{-1}$ Mpc--10 h$^{-1}$ kpc: galaxy formation,
correlations, and velocities)?

One variant of standard CDM that has received much
attention recently (e.g. Cen \etal 1992, Adams \etal
1993, Liddle \& Lyth 1992b) keeps all the usual assumptions
except the Zel'dovich primordial spectrum $|\delta_k|\propto
k^n$ with $n=1$, substituting instead ``tilted'' spectra
with $n\approx0.5-0.7$ which arise from more or less
complicated inflationary models.  Such models have the
virtue of being very well specified, with $n$ being the only
additional parameter beyond those of standard CDM. However,
it appears that ``tilted'' CDM is marginal at best.  For
example, for $n$ sufficiently small to account for the
observed large scale structure, there is probably too little
early galaxy formation.  Of course, it is possible to get
much more general non-Zel'dovich primordial fluctuation
spectra from inflation (see e.g. Primack 1991 for a review),
but these ``designer spectra'' are neither well motivated
nor well specified.

Cold + Hot Dark Matter (C+HDM) is physically at least as
well motivated as tilted CDM or any other variant of CDM
that we know.  Moreover it is well specified and has only
one additional parameter beyond those of standard CDM, the
neutrino mass or $\Omega_\nu$.  The currently available
solar neutrino data suggest that the electron neutrino and
one other neutrino, say the muon neutrino, have nonvanishing
mass, with $m(\nu_\mu)=(2-3)\times 10^{-3}$ eV. In simple
``seesaw'' models of neutrino masses in which the three
generations of heavy right-handed neutrinos are assumed
to be degenerate in mass, $m(\nu_\tau) \approx
0.3 (m_t/m_c)^2 m(\nu_\mu)$, where $m_t$ and $m_c$ are the top
and charm quark masses, so $m(\nu_\tau)$ should lie in the
relevant mass range $\sim10$ eV for C+HDM (for relevant
references and recent examples of such models, see Ellis,
Lopez, \& Nanopoulos 1992 and Dimopoulos, Hall, \& Raby
1992).  Moreover, the CHORUS and NOMAD $\nu_\mu \nu_\tau$
oscillation experiments now underway at CERN could see a
signal within a few years if these neutrino mass models are
right and the mixings are large enough.

While it is true that C+HDM does have an extra knob to
adjust compared to CDM, it also has a physical feature that
is suggested by the data: the neutrinos provide an
unclustered dark matter component on small scales, which
could help explain why dynamical estimates give $\Omega<1$
on small scales.  The out-of-equilibrium relativistic
Fermi-Dirac statistics of the neutrinos (once the neutrinos
decouple, their momenta just redshift; see e.g. Weinberg
1972, p. 535) enhances this effect, as we will discuss
below.

The main objection to C+HDM in principle is the apparent
unliklihood of having two different dark matter components
each making comparable contributions to the mass density.
Although one of the earliest C+HDM papers (Shafi \& Stecker
1984) proposed a particle physics model to account for this,
it must be admitted that we are unaware of any such model
that is attractive. However, the entire particle physics
Standard Model begs for further explanation, so it does not
disturb us to contemplate one more feature that, if valid,
would call for a more fundamental justification.

The question that we will consider here is whether C+HDM can
be valid: can it account for the astronomical data?  Basic
properties of mixed dark matter models were worked out some
time ago (Fang, Li, \& Xiang 1984, Valdarnini \& Bonometto 1985,
Achilli, Occhionero, \& Scaramella 1985); and the fact that C+HDM is a
promising model for large scale structure was established by several
previous linear calculations (Holtzman 1989, Schaefer, Shafi, \&
Stecker 1989, van Dalen \& Schaefer 1992, Schaefer \& Shafi 1992,
Taylor \& Rowan-Robinson 1992, and Holtzman \& Primack 1993).   A
simplified nonlinear calculation in a 14~Mpc box has been done by
Davis, Summers, \& Schlegel (1992), with the initial neutrino
fluctuations set equal to zero.  Gelb, Gradwohl, \& and Frieman (1993)
includes a cdm simulation in a 200~Mpc box starting from a linear
fluctuation spectrum inspired by C+HDM.  In the present paper we
present the first detailed nonlinear calculations for C+HDM of which we
are aware, with proper initial conditions, sufficiently many hot
particles to sample velocity space adequately, and a careful analysis
of dark matter and galaxy correlations and velocities with comparisons
to the available data.

This paper is organized as follows: \S 2 describes our
calculation of the cold and hot fluctuation spectra and
gives fitting functions, \S 3 explains the details of our
numerical techniques and spectrum normalization, and \S 4
explains our biasing scheme and galaxy finding algorithms.
The remaining sections present our results: \S 5 the
abundance of halos from our numerical simulations compared
to a Press-Schechter calculation, then the matter and galaxy
correlation functions (\S 6) and velocities (\S 7), and in
\S 8 the density distribution function.  In \S 9 we
summarize our results and the status of C+HDM as a model of
cosmological structure formation.  It fares remarkably well!

\bigskip
\centerline{\bf 2. The spectrum of fluctuations}

The initial spectrum of fluctuations for our C+HDM model is
computed using linear theory. The linear calculations trace
the evolution of fluctuations in five components: baryons,
photons, cold particles, massive hot particles, and massless
hot particles. The relative densities of baryons, cold
particles, and massive hot particles are here assumed to be
$\Omega_{\rm cdm}=0.6$, $\Omega_{\nu}=0.3$, $\Omega_b=0.1$,
using $h\equiv H_0/100\ \kms \Mpc\!^{-1}=0.5$ for the Hubble
parameter. The energy density of the photons is determined
from the observed mean temperature of the microwave
background, taken here to be 2.7 K. The massless hot
particles are identified with two species of massless
neutrinos, which determines the energy density in the
massless hot particles as compared with the density of the
radiation. The massive hot particle is taken to be a light
neutrino with a velocity distribution given by relativistic
Fermi-Dirac statistics.

For the initial conditions for the calculations, we have
assumed adiabatic perturbations with a scale-invariant
spectrum as predicted by standard theories of inflation. The
linear calculations are performed using the techniques of
Holtzman (1989). The coupled evolution of perturbations is
computed from very early times until the present. The
angular distribution of the radiation perturbations is
expanded in Legendre polynomials, and sufficient resolution
(up to $\sim 600$ orders at the time of recombination) is
kept so that small angle anisotropies can be calculated. The
momentum dependence of the massive neutrino perturbations is
determined by separately calculating the evolution of these
perturbations at 15 different comoving momenta. For the
massive neutrinos, only the lowest two angular moments of
the perturbations contribute gravitationally, so only these
moments are computed, using techniques similar to those
described by Bond and Szalay (1983).  This computation,
which involves the solution of integro-differential
equations, becomes lengthy at late times. However, once the
massive neutrinos become nonrelativistic, the high order
angular perturbations become small, and their subsequent
evolution can be computed with the full angular dependence,
which is described by a coupled set of differential
equations.  The switchover is made at the epoch $(1+z)\sim
220$, when the rest mass of the neutrino dominates the total
energy even for the highest momentum neutrinos. Additional
time is saved in computing the small scale perturbations by
setting the neutrino perturbations to zero at intermediate
times when they contribute less than 0.5 percent of the
total gravitational perturbation; once the neutrinos become
nonrelativistic and their fluctuations begin to grow, their
gravitational influence is turned on again.

A qualitative discussion of the linear fluctuation spectrum
is given by Holtzman (1989). In Figure~\nxtplt\PLOTA, we
present the computed spectrum for our cosmological model
along with some analytic fits (dashed curves).
Figure~\Plot$\!$a presents the fluctuations in the cold
particles, while Figure~\Plot$\!$b presents the fluctuations
in the neutrinos.  Note that even at the current epoch (top
curves in Figures~\Plot$\!$ab) the amplitude of the
neutrino fluctuations is different from that of the cold
particles; this arises because the neutrino perturbations
are damped by free streaming and Landau damping until the
massive neutrinos become nonrelativistic, and the neutrino
perturbations have not yet had enough time to grow to match
the fluctuations in the cold particles. The fluctuation
spectrum of the baryons quickly grows to match the spectrum
of the cold particles after recombination; by z=25, the
baryon fluctuations are within a few percent of the cold
particle fluctuations, and in our dissipationless
calculations we can treat them with sufficient accuracy by
including them with the cold fluctuations.

The power spectrum of total density
is $P(k,z)=(0.7\sqrt{P_{\rm cold}}+0.3\sqrt{P_{\nu}})^2~$.
The analytic fits shown in Figure~\Plot  are given by
$$\eqalign{
   P_{\rm cold}(k,z) &= {\ln(1+18k\sqrt{a}) \over
18\sqrt{a}}\cr
                   &\times\left[
                      1+1.2 k^{1/2} -27k +347(1-\sqrt{a}/5)k^{3/2}
                       -18(1-0.32a^2)k^2\right]^{-2}, \cr
  P_{\nu}(k) &= P_{\rm cold}(k,z){\exp(-q/16)\over 1 +0.03q +0.67q^2},
\hskip 2em q\equiv k/\sqrt{a},\cr}\nxt
$$\EQA
where $a=(1+z)^{-1}$ is the expansion parameter, and $k$
is the wavenumber measured in units of ${\rm Mpc}^{-1}$.
These approximations are valid for $k < 30{\rm Mpc}^{-1}$
and $z<25$. The accuracy (maximum deviation) of the $P_{\rm
cold}$ approximation is  5\%, while the neutrino spectrum is
accurate within 25\%.
(At $z=15$, when our simulations start, the maximum error in
the neutrino spectrum is 20\%, at $k\approx 0.06$
Mpc$^{-1}$; the next-largest error then is about 10\% at
$k\approx 0.3$.  Note that a 20\% error in the neutrino
spectrum leads to a 6\% error in the power spectrum of total
density, comparable to the maximum error in the cold
spectrum.) At high $k$, the leading term in the spectrum for
the ``cold'' component is the term with $k^{3/2}$, the last
term in the denominator providing a correction.

Wright \etal (1992) introduced the quantity ``Excess Power''
$$
 EP \equiv
3.4 {\sigma_\rho(8 h^{-1}\Mpc)\over \sigma_\rho(25
h^{-1}\Mpc)}\nxt
$$
as a measure of the shape of the spectrum, and pointed out
that $EP\approx1.30\pm0.15$ is required to fit the APM
$w(\theta)$.  (Here $\sigma_\rho(r)$ is as usual the rms
mass fluctuation in a top-hat sphere of radius $r$
calculated using $P(k,z)$.) $EP$ is defined so that $EP=1$
for standard CDM.  For our C+HDM model, $EP=1.37$,
consistent with the APM data.

\bigskip
\centerline{\bf 3. N-body Simulations}

\bigskip
\leftline{\it 3.1 Code}

Numerical simulations were done using standard Particle-Mesh
(PM) code (Hockney \& Eastwood 1981). The equations we
actually solved are given by Kates, Kotok, \& Klypin
(1991). Most of results are based on three simulations with
$256^3$ grid points of resolution of gravitational forces.
Three other simulations with lower resolution ($128^3$ grid
points) were mainly used for tests. Each simulation had
seven sets of $128^3$ particles. One set of particles
represented ``cold'' particles (WIMPs/axions plus baryons),
while the remaining six sets were used to simulate ``hot''
particles (neutrinos). The particles had different relative
masses: each ``cold'' particle had a relative mass 0.7 and
each ``hot'' particle had the mass $0.3/6=0.05$.

\bigskip
\leftline{\it 3.2 Thermal velocities of ``hot'' particles}

The six sets were arranged in three pairs, particles of each
pair having random ``thermal'' velocities of exactly equal
magnitude but pointing in opposite directions. The directions
of these ``thermal'' velocities were random. The magnitudes
of the velocities were drawn from relativistic Fermi-Dirac
statistics:
$$
        dn(v) \propto {v^2 dv \over \exp[v/v_0(z)]+1}, \nxt
$$
where $v_0(z) =(1+z) c kT_{\nu}/m_{\nu}c^2$, $c$ being the
velocity of light, $T_{\nu} = 1.95$~K, and
$$
m_{\nu}c^2 = 91.5\Omega_\nu h^2\ {\rm eV}
= 6.86(\Omega_\nu/0.3)(h/0.5)^2 {\rm eV}. \nxt
$$
Note that $v_0(0) =7.2(m_{\nu}/7{\rm eV})^{-1}\kms$ and that
the rms velocity $v_\rms(z)=3.596v_0(z)$. Thus at $z=10$,
for example, $v_\rms=290 \kms$, certainly high enough to
suppress $\nu$ clustering on galaxy scales.  These
relatively large velocities account for the falloff of
$\Delta$ at high $k$ in Figure~\Plot$\!$a.

The reason to have many particles to simulate the neutrino
distribution is that the motion of ``hot'' particles due
only to the thermal velocities must not generate any
spurious fluctuations.  The arrangement of hot particles in
pairs having oppositely directed velocities locally preserves
the center of mass, thus more closely simulating ``thermal''
velocities of ``hot'' particles. With the center of mass
preserved, the power spectrum of spurious fluctuations due
to the discreteness is restricted to have $k^4$ dependence
on the wavenumber $k$, which significantly reduces
propagation of errors to large scales, while the amplitude
at small scales is small thanks to the large number of
particles.

Test runs have been done to estimate the effects of the
discreteness. The test simulations had $32^3$ particles in
each set moving in a grid of $64^3$ cells. The size of the
box was chosen to be 14 Mpc. No initial fluctuations were
imposed on the particles. Thus ``cold'' particles initially
had no velocities and were placed in a regular cubic grid.
``Hot'' particles had thermal velocities, which due to the
discreteness effects produced fluctuations of density and
displacements of ``cold'' particles. Two simulations were
run: one with six sets of ``hot'' particles as described
above, another (like that of Davis, Summers, \& Schlegel
1992) with only one set of ``hot'' particles.  The latter
simulation did not preserve the local center of mass and
fluctuations were larger. We started both simulations at
redshift $1+z =20$. At the final moment, corresponding to
redshift zero, the level of density fluctuations was
$\sqrt{\xi(1)} =0.16$ for the simulation with six sets and
$\sqrt{\xi(1)} =0.38$ for the simulation with one set, where
$\xi(1)$ is the correlation function at one cell spacing
(0.22 Mpc). Thus the fluctuations in the six-set simulation
were satisfactorily small --- a factor $\sqrt{6}$ smaller
than in the one-set simulation. The correlation function at
one cell spacing is sensitive to very short scales. Another
statistic, which is sensitive to longer waves, is the
kinetic energy. In the absence of fluctuations, the kinetic
energy must decay as $a^{-2}$, where $a$ is the expansion
parameter. Because of spurious fluctuations induced by the
discreteness, the actual kinetic energy is larger than the
initial kinetic energy rescaled to the final moment.  The
ratio of the actual to the rescaled kinetic energies at
$z=0$ was $1.05$ for the six-set simulation and $1.43$ for
the one-set simulation. Visual inspection of dot plots of
``cold'' particles at the final moment confirmed our
conclusion that the simulation with six sets of ``hot''
particles mimics the neutrino thermal velocities with much
better accuracy.

\bigskip
\leftline{\it 3.3 Initial conditions and normalization}

 We used our analytical approximations (Equations
(\the\eqA)) for the ``cold'' and ``hot'' spectra. Initial
positions and velocities of particles were set using the
Zel'dovich (1970) approximation as was first described for
the three-dimensional case by Klypin \& Shandarin (1983). We
did not differentiate the velocity potential numerically as
Efstathiou \etal (1985) did, because it generates excessive
numerical noise at short scales. Instead, the displacement
vector was simulated directly. Phases of fluctuations were
exactly the same for ``hot'' and ``cold'' particles. When
generating velocities of ``hot'' particles, the  ``thermal''
component, as described above, was added to the velocity
produced by the Zel'dovich approximation.

The amplitude of fluctuations was normalized so our
realizations are drawn from an ensemble producing the
quadrupole in the angular fluctuations of the cosmic
microwave background at the $17\mu K$ level measured by COBE
(Smoot \etal 1992, whose power-spectrum-normalized
quadrupole actually includes only $l\ge3$ angular
components).  To be more specific,
we present expressions for the large scale angular
fluctuations, density fluctuations, and velocity
fluctuations. If $P_{\rm cold}(k)$ is the power spectrum of
density fluctuations of ``cold''  particles, then the
corresponding angular fluctuations $a_{lm}^2$, rms density
fluctuations $\sigma_{\rho}$, and rms 3D peculiar velocity
$\sigma_v$, are
$$
   a_{lm}^2 ={4 \over R_H^3}
             \int_0^{\infty}{dx\over x^3}P_{\rm cold}(k)J^2_{l+1/2}(x),
                \hskip 2em x=kR_H, \nxt
$$
$$
   \sigma^2_{\rho, {\rm cold}} ={1\over 2\pi^2}
             \int_{k_{\rm min}}^{k_{\rm max}}
                {dk k^2}P_{\rm cold}(k), \hskip 2em
   \sigma^2_{{v}, {\rm cold}} = {H^2\over 2\pi^2}
             \int_{k_{\rm min}}^{k_{\rm max}}
                {dk}P_{\rm cold}(k), \nxt
$$
where $R_H=6\times 10^3 h^{-1}$~Mpc is the radius of the
horizon, $k_{\rm min}$ and $k_{\rm max}$ are the limits
imposed by the finite size of the computational box, and
$J_n$ is the Bessel function. The rms angular fluctuations
due to all multipole components with given $l$ is then
$$
Q_l^2={(2l+1)\over 4\pi}a_{lm}^2
     ={2l+1\over \pi^2 l (l+1)}{A\over R_H^4}~,\nxt
$$
where $A$ is a normalization of the spectrum at small $k$:
$P_{\rm cold}=Ak$. In order to normalize the spectrum of
fluctuations to given rms of the quadrupole $Q_2$ we set the
normalization constant to
$$
   A = {6\pi^2 \over 5} Q_2^2R_H^4~.\nxt
$$
Note that the integral for the quadrupole $Q_2$ converges at
very long waves and this is the reason why we can use
$P_{\rm cold}$ instead of the spectrum of total
fluctuations: the hot and cold spectra are identical in the
long wavelength limit.

A number of authors (Liddle \& Lyth 1992a, Krauss \& White
1992, Davis \etal 1992, Adams \etal 1992, Lucchin,
Mattarese, and Mollerach 1992, and Souradeep \& Sahni 1992)
have recently reminded us that the CMB fluctuations observed
by COBE could include some contribution from tensor modes
(i.e., inflation-induced gravity waves, Starobinsky 1979) as
well as scalar modes (density fluctuations).  However, this
is a relatively small effect in many models of inflation,
comparable to the uncertainty in the COBE quadrupole
including the cosmic variance, and we will ignore it here.
Note that a larger tensor/scalar ratio in the large-angle
CMB fluctuations would imply a smaller amplitude for the
scalar fluctuation spectrum, and therefore a larger linear
bias factor for our C+HDM model.

To normalize the numerical simulations, we estimate
$\sigma_{\rho,{\rm cold}}$ and $\sigma_{v}$ expected in a
box of the same size $L$ as our computational box. To make
these estimates we set the limits of integration to $k_{\rm
min}=(2\pi/L)/\sqrt{2}$ and $k_{\rm max}=(2\pi/L) N/2$,
where $N$ is the number of particles along an axis (128 in
this case) and the factor $\sqrt{2}$ takes into account the
fact that each harmonic in the simulation represents a small
cubic element in the phase space. (The factor actually
depends on the shape of the spectrum.) Then we make a
realization that gives the same $\sigma_{v}$ as obtained
from numerical integration of the spectrum. Usually the
spectrum of density fluctuations is already accurate within
10--15\%. Finally we estimate the power spectrum of density
fluctuations in the numerical simulation and compare it with
the predicted one. A correction to the amplitude is made if
necessary.  The agreement of the power spectrum of density
fluctuations in the numerical simulation with the prediction
of the linear theory is the final normalization criterion.

\bigskip
\leftline{\it 3.4 Simulations}

 Sizes of the computational boxes for the C+HDM simulations
were 14 Mpc, 50 Mpc, and 200 Mpc (the Hubble constant is
assumed to be $H=50\kms \Mpc\!^{-1}$, i.e., $h=0.5$). The
simulations were started at redshift 15 and run to redshift
zero with a constant step $\Delta a=0.01$ in the expansion
parameter $a$. Each simulation had  $128^3$ ``cold''
particles and $6\times 128^3$ ``hot'' particles moving in a
grid of $256^3$ cells. Thus, the smallest resolved comoving
scale  was 0.055 Mpc, 0.195 Mpc, and 0.781 Mpc for the three
simulations, and correspondingly the mass of a ``cold''
particle was $6.29\times 10^7M_{\odot}$, $2.86\times
10^9M_{\odot}$ and $1.83\times 10^{11}M_{\odot}$.

As explained in the previous subsection, the simulations
were normalized to the COBE quadrupole $Q_2 =17\mu K$. This
normalization corresponds to a biasing parameter $b=1.5$,
where $1/b$ is defined as the rms fluctuations of mass
inside a sphere of radius $8h^{-1}\Mpc$. Another way of
expressing the normalization is to give the rms peculiar
velocity at some scale as predicted by the linear theory.
The total (i.e., with no smoothing) rms peculiar velocity of
``cold'' particles is $\sigma_v=750\kms$, and the rms
velocity of a sphere with radius $50h^{-1}$ Mpc is
$V_{50}=360\kms$. For comparison, CDM with the same biasing
parameter $b=1.5$ predicts $\sigma_v=660\kms$ and
$V_{50}=190\kms$. (We compare these linear $V_R$
predictions with the data in \S 9, below.)

A model with the CDM initial spectrum was also simulated. We
used the spectrum given by Bardeen \etal (1986). The spectrum
was normalized to give the same biasing parameter $b=1.5$
as for our C+HDM simulations. This normalization corresponds
to the rms of the quadrupole $Q_2=8.5 \mu$K, which is
not compatible with the COBE DMR results. The size of the
computational box was 50 Mpc. The initial phases of
fluctuations, initial moment, and time-step were exactly the
same as for the C+HDM simulation with the same box size.

Figure~\nxtplt\PLOTB presents the distribution of matter in
a rectangular slice of 10~Mpc thickness in the 200~Mpc box
simulation.  The slice was chosen to pass through the
highest density concentration in the simulation. The density
concentration is a rich galaxy cluster in the upper right
quadrant of the slice. Figure~\Plot$\!$a presents only a
small (8\%) random fraction of ``cold'' particles. The
distribution of ``cold'' points for which the density is
higher than $\rho_{\rm total} > 10\langle \rho _{\rm
total}\rangle$ is shown in Figure~\Plot$\!$b ($1/6$ of the
points). The distribution of 1015 ``galaxies'' in the slice
is shown in Figure~\Plot$\!$c. ``Galaxies'' were first
preselected as maxima of density, and then those with mass
larger than $10^{11}\Msun$ are plotted. The area of the
circle presenting a ``galaxy'' is proportional to its mass.
The distributions of ``hot'' and ``cold'' particles are
significantly different at small scales even at the present
time. Figure~\nxtplt\PLOTM shows the distribution of both
components in a 220~kpc slice in the 14~Mpc simulation. Note
that while all large condensations in the plot are seen both
in ``cold'' and ``hot'' particles (like the large group in
the left top corner, a few big ``galaxies'' in the filament,
and the filament itself), small halos of ``cold'' particles
do not find counterparts in ``hot'' particles (for example,
the two ``galaxies'' in the void at the middle right).

\bigskip
\centerline{\bf 4. Biasing schemes and galaxy finding algorithms}

In general, the distribution of galaxies (``luminous''
matter) does not follow that of the dark matter. Thus we say
that the galactic distribution is ``biased''. There are
different notions hidden behind the term ``bias''. The
simplest one is the linear biasing parameter $b$, which is
just the normalization of the spectrum of fluctuations. For
the C+HDM model normalized to the quadrupole $17\mu K$ the
biasing parameter is $b=1.5$, meaning that in a sphere of
radius $8h^{-1}\Mpc$ the level of density fluctuations as
estimated by the linear theory is 1.5 times smaller than the
level of fluctuations of the number of galaxies. Thus with
this large scale normalization, the C+HDM model must be
``biased.''  Another important notion is ``physical bias'':
luminosity density is not proportional to the total local
density because the efficiency of galaxy and star formation
can depend very nonlinearly on the density and can also
depend on other aspects of the environment. Note that this
could happen even for a model with linear biasing parameter $b=1$.
Physical bias is perhaps very complicated,
involving many poorly understood aspects of galaxy
formation.  Although CDM can perhaps be rendered more
consistent with the observed large scale galaxy distribution
and other data by invoking a physical bias mechanism
(e.g., Babul \& White 1991), this weakens the predictive
power of the CDM model which was in our opinion one of its
most attractive features.

There are some assumptions that can simplify the situation.
Thus if some particle never resided in a high density region,
it cannot be a part of a galaxy. On the other hand, if a
dense clump has the mass and the radius of a large galaxy,
it probably is a galaxy. This is not necessarily true for an
object the size of a dwarf galaxy, however, because
supernova explosions could have swept away most of its gas,
thus stopping star formation.

Various algorithms have been used to find ``galaxies'' in
numerical simulations. Cluster-analysis or ``friends-of-friends''
(Efstathiou \etal 1985, Frenk \etal 1988) has the advantage
that no specific shape of an object is assumed.
Unfortunately it also has significant disadvantages. If the
neighborhood radius (effectively a density threshold) is
small, it  does not find all halos (especially in low
density regions). If the radius is larger, then it does not
resolve halos in dense regions: in the case of an object
which looks like a cluster of galaxies (judging from its
radius and mass) the algorithm assigns all the mass of the
cluster to one ``galaxy'', thus mixing together galaxies,
groups, and clusters.

We decided to use simpler prescriptions. Three different
procedures were used to identify the ``luminous'' matter in
our simulations. All these procedures deal with the density
at a given moment of time. When possible, we compare results
obtained with different methods or with different parameters
in the same method.

\noindent
i) {\it Density thresholding}. If we do not need to deal
with separate objects (for example, in the case of the
correlation function), we find the density of ``cold''
particles on our standard $256^3$ mesh and consider a cell
``luminous'' if the density is larger than some threshold
$\rho_{\rm thr, cold}$. The luminosity density is assumed to
be directly proportional to the density at the cell. This
simple prescription removes essentially all points from low
density regions that have not yet collapsed. The density
threshold is the free parameter. For the 50~Mpc simulation
the cold density threshold was chosen to be $\rho_{\rm thr,
cold}=25\langle \rho_{\rm cold} \rangle$, which corresponds
to the mass in a cell larger than $8.9\times 10^9M_{\odot}$.
This threshold is a reasonable compromise between
expectations for the galactic mass, overdensity for
collapsed objects, and statistical fluctuations in the
correlation function. We discuss below the dependence of the
correlation function on the threshold. The situation is
worse for the 200 Mpc simulation because each ``cold''
point has enough mass to represent a galaxy. Nevertheless,
we chose a threshold $\rho_{\rm thr, cold}=5\langle
\rho_{\rm cold}\rangle$, thus selecting cells that collapsed
in the simulation. This corresponds to the mass in a cell
larger than $1.1\times 10^{11}M_{\odot}$. To be precise, the
thresholding of density does not guarantee that the object
had undergone collapse.  But in the case of the C+HDM
model the effective slope of the power spectrum on
galactic scales is about $-2.5$, which implies that the
formation of an object goes first through the formation of
pancakes (e.g., Klypin \& Melott 1992). The Zel'dovich
approximation for pancake formation shows that the points
with density at the threshold $5\langle \rho_{\rm cold}
\rangle$ will collapse very soon.

\noindent
ii) {\it Maxima of density}. If we need to deal with
separate objects (mass function, pairwise velocities), we
find maxima of density. The total (``cold'' plus ``hot'')
density was produced on the $256^3$  grid. Then all local
maxima above some threshold $\rho_{\rm thr}$ are found. The
way we assign mass to a maximum (``galaxy'') depends on the
particular problem we deal with. For counts of objects at
different redshifts the mass of a ``galaxy'' is the mass in
the cube of $3\times 3\times 3$ cells centered on the
maximum. Effectively this corresponds to the radius of
objects being 1.5 cell units. For simulations with 14~Mpc and
50~Mpc box sizes, this corresponds to 82~kpc and 293~kpc.
For all the simulations the threshold of the density was
chosen to be $\rho_{\rm thr}= 50\langle \rho_{\rm
total}\rangle$. The Press-Schecter (1974) approximation and
visual investigation of dot plots were used as guides for
the choice of the parameters: with smaller size of the
template we significantly underestimate the mass of halos in
a dense environment, while with lower density threshold we
overcount objects as compared with the Press-Schecter
approximation.

\noindent
iii) {\it Maxima of density inside a sphere}. For the
analysis of peculiar velocities of ``galaxies'' we do not
need to know the mass of a ``galaxy'' precisely, but we
would like to know its position slightly more accurately.
After locating positions of maxima as for method (ii), we
place a sphere of radius $r$ at the position of each maximum
and find the center of mass of all ``cold'' particles inside
the sphere. We then displace the center of the sphere to the
center of mass and repeat the procedure. After two
iterations the positions of the spheres essentially stop
changing. The typical distance between the final and initial
centers of a sphere is less than $1/4$ of the cell size, and
the increase in the number of points inside a sphere is 20--40\%.
One can expect that this procedure gives a slight improvement
over biasing scheme (ii). The typical radius of a sphere
was 0.5 of the cell size.

\bigskip
\centerline{\bf 5. Abundance of halos}

In order to estimate the number density of galaxies at
various redshifts in our simulations we found all local
maxima of the total density above the overdensity 50 (\S4,
prescription (ii)). Figure~\nxtplt\PLOTC shows the number
density $N(>M)$ of dark halos with mass larger than $M$.  The
three bottom full curves (mass limits $M=3\times 10^{11}\Msun$
and higher) are for the simulation with 50~Mpc box size. The
three top full curves correspond to the number of objects in
the simulation with 14~Mpc box size. The dot-dashed curves
show results for the Press-Schecter approximation, which
were estimated as follows. For the Gaussian filter with
radius $R_f$ and mass $M=(2\pi)^{3/2} \rho_0 R_f^3$ the
number density of halos with mass larger than $M$ is
$$
  N(>M,z)=\int_M^{\infty} n(m,z) dm
         ={\delta_c\over 2\pi^2}\int_{R_f}^{\infty}
          {\epsilon(z)\over\sigma(r_f,z)}
          \exp\left({-\delta_c^2\over 2\sigma(r_f,z)^2}\right)
          {dr_f\over r_f^2}~,\nxt
$$
where
$$
    \epsilon(z) ={\int k^4 P(k,z) e^{-(kr_f)^2} dk} {\bigrm /}
                  \int k^2 P(k,z) e^{-(kr_f)^2} dk~,\nxt
$$
and $P(k,z)=(0.7\sqrt{P_{\rm cold}}+0.3\sqrt{P_{\nu}})^2~$
is as usual the power spectrum of the total density. The
parameter $\delta_c$ was considered as a free parameter. The
best results for 14~Mpc and 50~Mpc boxes and for the CDM
model simulated in a 50~Mpc box were found for
$\delta_c=1.60$, which is very close to the value given by
the top-hat model.

As one could expect, because long waves are not present in a
finite box and because high peaks are rare in the box, the
number of massive objects at high redshifts in the numerical
simulations was below predictions of the Press-Schecter
approximation. Although (as we explained in \S4) the
parameters of our galaxy finding algorithm were chosen to
have results consistent with the Press-Schecter
approximation, the fact that we succeeded for such a wide
interval both in mass and redshift should be considered as a
remarkable success for the approximation (cf. Efstathiou \&
Rees 1988, Bond \etal 1992).

Figure~\nxtplt\PLOTD\ shows the number density of halos per
unit volume and per unit logarithmic mass interval at
redshift zero. The two full curves present the number of halos
in C+HDM simulations for 50~Mpc box size (right curve) and for
the 14~Mpc simulation (left curve). Results for the CDM simulation are
shown by the dash-dotted curve. The dashed curve presents an
analytical approximation:
$$
 {dn\over d\log{M}} =3.1\times 10^{-3}\left({M\over M_*}\right)^{-1+1/12}
                     \exp{-\left({M\over M_*}\right)^{1/6}},
        \hskip 2em M_* =2\times 10^{12}\Msun, \nxt
$$\EQB
The power law slopes in this approximation were taken from
the Press-Schecter approximation for a slope $n=-2.5$ in the
power spectrum, which is appropriate for the C+HDM model. We
did not estimate the amplitude and the characteristic mass
scale $M_*$ using the P-S approximation, so Equation (\the\eqB)
should be considered as a fit to the simulation data.

It is interesting to compare the numbers of ``galaxies'' in
the C+HDM and in the CDM simulations. The number density of
``galaxies'' in the CDM simulations at different redshifts is
shown in Figure~\nxtplt\PLOTO\ (triangles are for the
numerical results and dashed curves are for the
Press-Schecter approximation with the parameter
$\delta_c=1.60$). Full curves are the results of the
Press-Schecter approximation for the C+HDM model. At
redshift $z=0$ the CDM model predicts about twice the number
of dark halos for masses $M=10^{11}\Msun - 10^{14}\Msun$.
For less massive objects both theories predict essentially
the same number of galaxies. The only masses where the C+HDM
beats the CDM model are those of rich galaxy clusters
$M>10^{15}\Msun$. At higher redshifts the C+HDM model
predicts progressively smaller number densities of objects.

All these curves are for CDM and C+HDM normalized with the
same linear bias factor $b=1.5$.  As is well known (Davis
\etal 1985), CDM with this low a bias factor predicts
galaxy peculiar velocities that are too large on small
scales; as we will discuss shortly, we confirm this in our
CDM simulations.  CDM with $b=2.5$ gives galaxy number
densities that are more like our $b=1.5$ C+HDM model.  This
is illustrated by the dot-dashed curve on the Figure.

We note that the C+HDM model predicts quite substantial
numbers of massive galaxies at high redshifts. For example,
at redshift $z=3$ each cube of a size 100~Mpc contains
$\approx 180$ galaxies with mass larger than $10^{11}\Msun$.
Unfortunately, it is difficult to make a conclusive
comparison with observations because there is still no
reliable estimate of the number of galaxies at high redshifts
with any given minimum estimated mass or circular velocity.
The number density of QSOs is another test for the
theory (Efstathiou \& Rees 1988); this has recently been
considered in a linear calculation as a possible constraint
on C+HDM (Schaefer and Shafi 1993). At redshift $z=2$ the
number of QSOs brighter than $L=10^{47}$ erg s$^{-1}$ is
$N_Q\approx 10^{-7}$ QSOs per cubic Mpc ($h=0.5$). If we
assume that galaxies with mass larger than $3\times
10^{11}\Msun$ are responsible for the QSOs, then in the
C+HDM model the fraction of galaxies hosting a QSO is about
$2.5\times 10^{-4}$ and each of the host galaxies deposited
about one per cent of its mass into the central mass
associated with the QSO. Thus, the C+HDM model probably is
compatible with the observed number of QSOs and galaxies at
redshifts up to $z=2-3$. But the C+HDM model will have
difficulties if significant numbers of massive galaxies or
bright QSOs are found at $z > (4-5)$. For example, at $z=5$
the number of halos in the C+HDM model with mass larger than
$10^{11}\Msun$ is $10^{-7}\Mpc^{-3}$. Thus if the number of
QSOs per cubic Mpc remains $N_Q\approx 10^{-7}$ up to $z=5$
(these QSOs could then plausibly generate enough UV
radiation to account for the Gunn-Peterson test and the
proximity effect, see Madau 1992), then every $10^{11}\Msun$
dark C+HDM halo should have a QSO inside.

Estimates of the number of very rare high redshift objects
based on the Press-Schecter approximation should be treated
with caution, however.  It is not clear whether the
approximation provides a good fit for very rare objects.
And even in the framework of the Press-Schecter
approximation there is a possibility to tune parameters
to give many more dark halos at high redshifts. The
parameter $\delta_c$, here assumed to be 1.60, could be
smaller at higher $z$ giving more objects. For example,
Efstathiou \& Rees (1988) used $\delta_c=1.33$. At $z=4.5$
the C+HDM model with $\delta_c=1.33$ predicts $N(>3
\times10^{11})=2.5 \times10^{-6}$, which is probably enough
to account for observed high redshift QSOs.

\bigskip
\centerline{\bf 6. Correlation functions}

The evolution of the correlation function of the total
density in the simulation with 50~Mpc box size is shown in
Figure~\nxtplt\PLOTF. The curves in the plot are labeled by
the expansion parameter. The correlation function $\xi(r)$
evolves much as in the HDM model: $\xi(r)$ is well
approximated by a power law, with the slope increasing with
time. At the final moment the slope of the correlation
function is about $\gamma = 1.8$, which is the observed
slope of the galaxy correlation function, but the
correlation length is $2.2 h^{-1}\Mpc$ and is too small as
compared to the correlation length of galaxies.  The dashed
curve in this figure shows the correlation function of the
``cold'' matter at the same epoch.  It is close to the
correlation function of the total density, which indicates
that there is not much difference between the distribution
of ``hot'' and ``cold'' particles.  For comparison we also
present the correlation function of the dark matter in the
CDM simulation. It is higher than that of the C+HDM
simulations and it does not behave as much like a power law.

The correlation length of the total matter is even smaller
for the 14~Mpc box: $0.82 h^{-1}$ Mpc. It is  slightly
larger in the 200~Mpc box simulation (Figure~\nxtplt\PLOTE).
The length is $2.9 h^{-1}$ Mpc for this case, but it is
still too small. The small correlation length of the dark as
well as the ``cold'' matter indicates that the C+HDM model
needs a bias to be consistent with the observed correlation
function of galaxies. Normalizing to the COBE quadrupole
led to the biasing parameter $b=1.5$. If we assume that the
correlation length for galaxies is $5.5 h^{-1}\Mpc$ and take
$2.9 h^{-1}\Mpc$ as an estimate of the correlation length in
the C+HDM model, than the biasing parameter as found from
the numerical simulations is $b=\sqrt{(5.5/2.9)^{1.8}}=1.8$,
which is reasonably close to the COBE normalization biasing
parameter. The actual biasing parameter for simulated
``luminous'' matter will be slightly higher ($b=1.9$), but
still in reasonable agreement with the COBE normalization.

To estimate the correlation function of ``luminous'' matter
in the model, we use biasing algorithm (i) of \S 4
(density thresholding). Figure~\Plot shows the correlation
functions of ``galaxies'' for the 50~Mpc C+HDM simulation
(dashed curves, mass in each selected cell is larger than
$8.9\times 10^9M_{\odot}$) and the 200~Mpc simulation (full
curves, mass in a cell is larger than $1.1\times
10^{11}M_{\odot}$). For comparison the correlation
functions of the ``cold'' matter are also presented (two
bottom curves). The dot-dashed line is the power law $\xi
=(5.5 h^{-1}/r)^{1.8}$, which gives a good approximation to
the ``galaxy'' correlation function from 200~kpc to 20~Mpc.

The correlation function of ``galaxies'' depends on the
biasing prescription. We found as expected that the
correlation function is increased if regions with higher
density or if more massive clumps are selected. The
influence of different effects on the correlation function
is demonstrated in Figure~\nxtplt\PLOTL. To estimate the
correlation functions presented in the plot, we have not
used the usual FFT technique, but instead we counted pairs
of objects as a function of the distance between them. This
accounts for small variations compared to previous plots.
The pairs of ``galaxies'' were weighted by the product of
the masses of the objects. This is needed in order to
compensate the ``overmerging'' effect in numerical
simulations: because of lack of resolution and because no
energy dissipation takes place when ``galaxies'' form,
substructures disappear when objects of smaller size merge
to produce a larger object.

The two full curves in Figure~\Plot$\!$a show how the
correlation function scales with the density threshold in
the 50~Mpc box simulation. The lower curve is for the
density threshold 10 and the upper curve is for the density
threshold 25. The dashed and dot-dashed curves show results
for ``galaxies'' found as maxima of number of ``cold''
particles inside a sphere of 100~kpc radius in the 50~Mpc
box simulation. The correlation function of ``galaxies''
more massive than $3\times 10^{10}\Msun$ (913 objects) is
shown as the dashed curve. The dot-dashed curve is for the
87 ``galaxies'' with mass larger than $3\times
10^{11}\Msun$. On scales below $\sim 1 h^{-1}\Mpc$, the
variations in the ``galaxy'' correlation function are about
a factor of two, the slope and the amplitude being higher
for objects that are denser or more massive. On larger
scales all curves are close to each other and the
correlation length essentially does not depend on particular
details of the biasing prescription and is equal to $(4\pm
0.5) h^{-1}\Mpc$. For comparison we show results for
galaxies in the CfA catalog (Davis \& Peebles 1983). Although
at small scales there is a reasonable agreement,
``galaxies'' in the simulation are systematically less
correlated than in observations. The lack of long waves in
the numerical simulation mainly accounts for the difference.
If the long waves are included (as in the 200 Mpc
simulation), the correlation length for ``galaxies'' rises
to the level $(5-6)h^{-1}\Mpc$.

Figure~\Plot$\!$b shows effects of peculiar motions on the
galaxy correlation function. Full curves in the plot present
correlation functions in redshift space (peculiar velocity
along the line of sight is interpreted as the additional
displacement) for 50~Mpc (bottom curve) and 200~Mpc (top
curve) simulations. Corresponding correlation functions of
913 and 2831 ``galaxies'' in real space are shown as dashed
curves. ``Galaxies'' in the 200~Mpc simulation were found as
maxima of the number of ``cold'' particles inside a sphere
of 300~kpc. The peculiar motions have two effects.
Correlations at small scales are reduced because large
velocities inside virialized groups spread the groups along
the line of sight. This affects scales up to $\sim 2 h^{-
1}\Mpc$, which corresponds to peculiar velocities of
$\approx 200\kms$. The large-scale motions increase
correlations, but the effect is small in our case. The
markers on the plot present the correlation function in
redshift space for galaxies in the CfA slices (Vogeley \etal
1992).

\bigskip
\centerline{\bf 7. Velocities}

One of the greatest difficulties for the CDM model is
getting the velocities right on small scales. For the
``unbiased'' (i.e., $b=1$) CDM model the peculiar velocities
were much too large (Davis \etal 1985) compared to observed
velocities in pairs of galaxies (Davis \& Peebles 1983).
Although ``velocity bias'' (Carlberg \etal 1990, Couchman \&
Carlberg 1992) perhaps reduces the pairwise velocities, it
apparently does not entirely eliminate the contradiction
with observations. Moreover, the amount of velocity bias, if
any, apparently depends on the simulation technique and
galaxy-finding prescription.  As we discuss at the end of
this section, our simulations do not support the suggestion
(Carlberg \etal 1990) that dynamical friction leads to
velocity bias.

For the ``biased'' CDM model of Efstathiou \etal (1985) with
biasing parameter $b=2.5$, the magnitude of the velocities
was about correct, but the dependence of the velocities on
the distance between galaxies was wrong: while the observed
pairwise velocities slowly rise with the projected distance
$\sigma=(340\pm 40)(hr_{\rm Mpc})^{0.13\pm 0.04}\kms$
(Davis \& Peebles 1983), relative velocities of pairs of
``galaxies'' in the model decreased with distance for $r <
1$ Mpc and only then started to rise.

In the C+HDM model the total rms velocity of matter relative
to the rest frame is $\sigma_v= 750\kms$ (the thermal
velocities are not included), which is larger than
$\sigma_v= 660\kms$ in the CDM model normalized to the same
biasing parameter $b=1.5$. However, most of the velocity in
the C+HDM model is due to the long waves. For example, the
rms velocity of a sphere with radius $50h^{-1}\Mpc$, which
is the characteristic scale of the Great Attractor, for the
C+HDM model is $V_{50}=360\kms$ compared to the CDM value
$V_{50}=190\kms$. Because of nonlinear effects peculiar
velocities in the numerical models were higher, but not much
higher, than the linear theory predictions. For the
50\Mpc box simulation the linear theory predicts the
rms 3D velocity $326\kms$, while in the simulation it was
$384.6\kms$. The 200\Mpc box simulation had larger
velocities: $\sigma_v=623\kms$ was predicted by the linear
theory and almost the same value ($617\kms$) was actually
found in the numerical simulations.

The distribution of relative peculiar velocities of pairs
is described by several functions. The first
moment $\langle v_{21}\rangle$ is the mean peculiar velocity
along the line joining the pair (a positive value means that
the two points move apart from each other faster than the
Hubble flow). The second central moment of the distribution
of pair velocities has two components: the velocity
dispersion along the separation vector ($v^2_{\parallel}$)
and the velocity dispersion perpendicular to the vector
($v^2_{\perp}$). Figures~\nxtplt\PLOTG~and
\nxtplt\PLOTI~present these components of the pair-wise
velocity distribution in the 50\Mpc box simulations for both
C+HDM and CDM models. The top panels show $\langle
v_{21}\rangle$ for ``cold'' particles (dashed curves) and
for ``galaxies'' (full curves). The dot-dashed curves stand
for the velocity equal to the Hubble velocity: a pair on the
curve would have zero {\it proper} velocity, thus indicating
a stationary pair in a statistical sense. The bottom panels
present the rms values of $v_{\parallel}$ (full curves) and
$v_{\perp}$ (dashed curves) for ``cold'' particles (the two
top curves) and for ``galaxies'' (the two bottom curves).
Because of the lack of resolution at small scales, one
expects that the velocities in the simulation are smaller
than they would be in reality. But the resolution (one cell
size) in the simulation corresponds to $97h^{-1}$~kpc, so
scales above 200~kpc--300~kpc were well resolved.

The ``galaxies'' were found using algorithm (iii) of
\S 4 (density maxima on the $256^3$ grid, then maxima of
numbers of ``cold'' points inside a sphere of 100~kpc
radius). The total number of ``galaxies'' was 913 in the
50~Mpc C+HDM simulation, and 1758 for the CDM simulation. The
``galaxies'' were more massive than $3\times10^{10}\Msun$.
For the C+HDM simulation the rms peculiar velocity of the
``galaxies'' relative to the rest frame was $340\kms$, which
is only slightly smaller than the velocity of all ``cold''
particles ($384.6\kms$). But the pairwise velocity of
``galaxies'' differs significantly (by a factor 1.8 at
$1h^{-1}\Mpc$) from that of the dark matter at distances
smaller than $3h^{-1}\Mpc$, while they are almost the same
at larger scales. For the CDM simulation the situation is
rather different. There is a larger difference between the
rms velocities of ``galaxies'' and the dark matter relative
to the rest frame ($416\kms$ for galaxies and $582.5\kms$
for the dark matter).  The difference of pair-wise
velocities also is larger at $1h^{-1}\Mpc$ (by a factor of
two), but at larger distances the difference does not go
away. Even at $7h^{-1}\Mpc$ the ratio of velocities is about
1.3.

While $\langle v_{21}\rangle$ for ``cold'' particles in the
C+HDM model shows that evolution still goes on in this
population (slight outflow at scales $<1$ Mpc and
inflow at (1--2) Mpc), the pairs of ``galaxies'' at scales
less than $\approx 2h^{-1}\Mpc$ look like stationary objects
with their peculiar velocities cancelling the Hubble
expansion. The same is true for the CDM model.

In order to obtain pair-wise projected velocities, we place
two ``observers'' at opposite corners of the computational
box and simulate line-of-sight velocities $V_p$ of
``galaxies'' by adding the line-of-sight component of the
peculiar velocity to the distance scaled by the Hubble
constant. The projected distance $r_p$ between a pair of
objects is estimated in the usual way as
$r_p=\tan(\theta_{12}/2)\,(V_{p1}+V_{p2})/H$, where $H$
is the Hubble constant and $\theta_{12}$ is the angle
between the objects. The distribution of velocity
differences $V_{p1}-V_{p2}$ is then studied as a function of
$r_p$.  Following Davis \& Peebles (1983), we do not include
pairs with $\Delta V > 1000\kms$ in the analysis.  The solid
curves in Figure~\nxtplt\PLOTH represent the velocity
dispersion of ``galaxies'' along line of sight as a function
of projected distance for the 50~Mpc box simulations for
both the C+HDM and the CDM models. The dashed curves in
the plot are for all ``cold'' particles and the dot-dashed
curves show the law $340 (hr)^{0.13}\kms$, which is a good
fit to the observations.  The C+HDM model produces projected
pair-wise velocities that give a very nice fit --- both in
amplitude and in the trend with the distance --- to the
observed velocities. The CDM model with the same bias
factor ($b=1.5$) gives a systematically larger amplitude by
a factor 1.2 -- 1.3.

The dependence of the velocity dispersion on different
effects in demonstrated in Figure~\nxtplt\PLOTJ. The top
panel shows results for the 200~Mpc box simulation. The
velocity dispersion is higher in this case, but in general
the trends are the same and the difference in the amplitude
is not too large. The fitting law is $380 (hr)^{0.13}\kms$,
which is still compatible with the results of Davis \&
Peebles (1983). We found no dependence on the mass of
galaxies, but with the available statistics the range
in mass was not large. The long-dashed curve in the bottom
panel gives results for 481 ``galaxies'' with mass larger
than $10^{11}\Msun$. If instead of cutting velocity
differences at $1000\kms$ we consider only pairs with the
relative distance smaller than $10 h^{-1}\Mpc$, the
dispersion rises by only 10 per cent (short-dashed curve).
The dispersion significantly grows if the cutoff in the
velocity difference is increased to $1500\kms$ (dot-dashed
curve). The same effect was found in the CfA catalog (Davis
\& Peebles 1983).

The difference in the velocities of the dark matter
particles and ``galaxies'' is called the velocity bias. We
find that the global velocity bias measured as the ratio of
rms velocities of all points to the rms velocities of
``galaxies'' is quite small --- 1.13 for the 50~Mpc box
simulation and 1.05 for the 200~Mpc simulation. We also have
shown that at scales larger than $5 h^{-1}$ Mpc, the
velocity distribution of ``galaxies'' approaches that of
``cold'' particles. The difference appears on small scales,
where velocities of ``galaxies'' are about half those of
``cold'' particles. The difference becomes even larger if we
select those ``cold'' particles that reside in high density
regions. Particles with $\rho > 25 \langle \rho_{\rm
mean}\rangle$ have velocities that on average are a factor
1.2 higher than the average for all ``cold'' particles.

One source of the bias is the internal random velocities
inside ``galaxies''. The velocities are present in the
analysis of ``cold'' particles, but they are removed from
the velocities of ``galaxies''. In our simulations internal
velocities  account for about half of the effect. Carlberg
\etal (1990) suggested that dynamical friction is the
main cause of the velocity bias they found. Because the
acceleration due to the friction is proportional to the mass
of an object, more massive objects will tend to sink to the
bottom of the gravitational well and move slower. Thus, one
expects that if dynamical friction were important, more
massive ``galaxies'' would move slower than less massive
ones. Figure~\nxtplt\PLOTK shows the distribution of the
total velocity of galaxies relative to the rest frame in the
simulation with 50~Mpc box (913 objects). The curve on the
plot presents the averaged rms velocity. Although the
circular velocities of ``galaxies'' range from $50\kms$ to
$500\kms$, there is no indication that the velocity
decreases with mass. (Note that with our force resolution in
this simulation of 0.2 Mpc and mass resolution of about
$3\times10^9 \Msun$, overmerging (see e.g. Gelb 1992) is not
a serious problem for our galaxy velocity calculations.  We
have made plots of the density distribution in some of our
groups and clusters, and checked that our galaxy-finding
algorithm finds many ``galaxies'' there.
Recent CDM calculations that also find no evidence for velocity
bias include dissipationless simulations by Ueda, Itoh, \&
Suto 1993, and hydrodynamic simulations by Katz, Hernquist,
\& Weinberg 1992.)

\bigskip
\centerline{\bf 8. Density distribution function}

The density distribution function is an indicator of the
dynamical state of a system. In the linear regime the
distribution function is a gaussian function. But even for a
relatively small amplitude of the fluctuations, the
distribution function significantly deviates from the
gaussian law at large densities because rare high
fluctuations become nonlinear and produce a small number of
high density objects that cannot be found in the gaussian
field.

There are two asymptotics for the evolution of the density
distribution function. For a model dominated by caustics
(like the HDM model) the distribution function was found by
Kofman (1991). The distribution function $P(\rho)$, which we
define as a probability to find a volume element with the
density $\rho$, has a tail $P(\rho)\propto\rho^{-3}$ for the
pancake approximation. Another interesting case is a
hierarchy of $N-$body correlation functions (Balian \&
Schaeffer 1988), which probably arises at an advanced
nonlinear stage. If the correlation functions scale as
$\xi_N(r_1,\dots,r_N)=\lambda^{\gamma(N-1)}\xi_N(\lambda
r_1,\dots,\lambda r_N)$, then the probability to find $N$
points in a volume $l^3$ is $P_N\propto N^{-2+\omega},
0\le\omega\le 1$. It was shown that for large $N$ the
probability $P_N$ has the scaling form $P_N\approx
h(N/N_c)/(\bar{\xi}N_c)$, where $\bar{\xi}$ is the average
correlation function in the volume $l^3$ and
$N_c=nl^3\bar{\xi}$, where $n$ is the mean number density.
The function $h(x)$ decreases exponentially for large $x$
and behaves as $x^{-2+\omega}$ for small $x$. The hierarchy
of the correlation functions reasonably agrees with
numerical results for the CDM model (Bouchet, Schaeffer, \&
Davis 1992).

Figure~\nxtplt\PLOTN shows the density distribution function
multiplied by density ($\rho dn/d\rho$) for the 50~Mpc box
C+HDM simulation at different redshifts. The distribution of
the total density (both ``cold'' and ``hot'' components
included) was obtained on $256^3$ mesh using the Cloud-In-Cell
algorithm. We do not show the function at small densities
because discreteness effects are too large. Figure~\Plot
presents values of $\sum_i{\rho_i(\rho;
\rho+\Delta\rho)}/\Delta\rho N, N=256^3$, which is an
estimator for $\rho P(\rho)$. The full curves are labeled by
the expansion parameter $a$. The function $P(\rho)$ grows
with time and its slope becomes smaller. By redshift $z=1$
the function has a power-law shape with the slope $\rho
P(\rho)\propto \rho^{-2}$, which is shown as the dashed
line. This is consistent with a large fraction of
matter being in pancake-type structures. The total density
contrast (as estimated by the linear theory) at that moment
was $(\delta\rho/\rho)_{\rm total}= 1.17$, which is the
amplitude characteristic for the formation of filaments.

At the final moment $z=0$ the density distribution showed
the power law behavior at small densities with the slope
$\approx -2.3$ and an exponential fall off at large
densities. The dot-dashed curve in Figure~\Plot shows the
fit to the distribution function
$$
   P(\rho)=0.275\left({\rho\over \langle\rho\rangle}\right)^{-2.25}
            \exp\left(-{\rho\over 2300\langle\rho\rangle}\right),
        \hskip 4em {\rho > 10 \langle\rho\rangle}.\nxt
$$\EQE
Let us compare with the results for the CDM model: Bouchet,
Schaeffer, \& Davis (1992) present counts of particles in
cubic cells for a CDM simulation normalized with biasing
parameter $b=2.5$. The size of their computational box was
64~Mpc. If we take the CDM results for the size of the cubic
cell having on average the same number of particles ($1/8$)
as a cell in our model, then the slopes of the distribution
function are the same ($-2.3$) in both models. Bouchet,
Schaeffer, \& Davis also give an approximation for the
function $h(N/N_c)$ introduced above: $h(x) \approx
0.075x^{-2}\exp(-0.08x)$. For our C+HDM simulation
$N_c=\bar\xi\approx 105$, $N=\rho/\langle\rho\rangle$, and
$h(x)\approx 0.086x^{-2.25} \exp(-0.046x)$, which is not
much different from the results for the CDM simulation.

\bigskip
\centerline{\bf 9. Conclusions and Discussion}

Let us begin by summarizing the conclusions we have reached
above from our C+HDM high-resolution PM simulations in 14,
50, and 200 Mpc boxes (always with $H_0=50$ km s$^{-1}$
Mpc$^{-1}$), our CDM comparison simulation in a 50 Mpc box,
and our Press-Schechter calculations.

\noindent 1) The correlation function of the dark matter
evolves as a power law, with slope increasing with time. At
redshift $z=0$, the slope is $\gamma=1.8$ and the
correlation length is $3h^{-1}\Mpc$. The correlation length
for ``galaxies'' in the model is $(5-6)h^{-1}\Mpc$ and it is
almost independent of the parameters of the biasing scheme.
The ``galaxy'' correlation function in redshift space
matches that of the CfA slices (Vogeley \etal 1992)
reasonably accurately.

\noindent 2) Pair-wise velocities of galaxies in the C+HDM
model are in a good agreement with the results for the CfA
catalog (Davis \& Peebles 1983): projected velocity
dispersion slightly increases with separation and its
amplitude at the distance 1~Mpc is $(340-380)\kms$. This
agreement must be regarded as an important success of the
model.

\noindent 3) At $z=0$ the mass distribution of dark halos is
roughly approximated by $dn/d\log M\propto M^{-1}$, which
is close to what was found for the CDM model, but the
amplitude of this approximation is about a factor of two
lower for the C+HDM as compared to the CDM model with the
same biasing parameter. The Press-Schecter approximation
(Equations 7, 8) provides a reasonably good fit for the
number of dark halos at all redshifts. At higher redshifts
the C+HDM model predicts a progressively smaller number of
dark halos of galactic mass. By redshift $z=2$ the
difference is about a factor of ten. We note that the C+HDM
model predicts roughly the same number density of galactic
halos as the CDM model with the biasing parameter $b=2.5$.
Although the C+HDM model does not contradict the number of
galaxies and quasars at $z< (2-3)$, more observational data
will be needed to test the theory at higher redshifts.

\noindent 4) The density distribution function at redshift
$z=0$ behaves as a power law with an exponential falloff.
The distribution agrees well with that for the CDM model and
with the predictions for the hierarchy of $N-$body
correlation functions of Balian \& Schaeffer (1988).

It was already known from the earlier linear C+HDM
calculations listed in \S 1 that C+HDM provides a good
fit to the available large scale structure data such as
galaxy large-angle correlations and rich galaxy cluster
spatial correlations.  Figure~\nxtplt\PLOTP shows the
remarkable agreement between the C+HDM predicted bulk flow
velocity as a function of radius and the POTENT
analysis of all the peculiar velocity data now available
(Dekel 1992).  That the same model can also account
for the few-Mpc-scale velocity data is a major triumph for
C+HDM, and reflects the special feature of this model: the
fact that its hot component is less clustered on small
scales than the cold and baryonic components.

Ostriker and Suto (1990) proposed to evaluate models on the
basis of a ratio of the streaming velocity to the random
velocity dispersion of galaxies, which they dubbed the
``cosmic Mach number'' ${\cal M}$. This ratio has the virtue
of being independent of the bias factor (i.e. spectrum
normalization), to first order. Unfortunately, the available
data do not yet permit a reliable estimate of ${\cal M}$,
but for guidance we will use ${\cal M}=2.2\pm 0.5$ for
$r_{\rm obs}=14.2h^{-1}\Mpc$ (Groth, Juszkiewicz, \&
Ostriker 1989).  Gaussian smoothing with $r_s=5h^{-1}\Mpc$
was used to remove small-scale nonlinear velocities. The
sample radius $r_{\rm obs}=14.2h^{-1}\Mpc$ approximately
corresponds to the size of a cubical volume $L_{\rm
obs}=58\Mpc$ (we use $h=0.5$ as usual, estimate the radius
of a top-hat window function and equate the volumes in the
cubical and spherical windows; in the numerical simulations
we estimate the Mach number using a cubic real-space window
function). For comparison, we use linear theory to estimate
the rms velocity of matter inside a volume of size  $L_{\rm
obs}=58$ Mpc. To find the noise velocity below $58\Mpc$ we
use the linear theory to get the contribution from
wavelengths between 58 Mpc and 50 Mpc and we use the
numerical simulations to estimate the rms noise velocities
due to waves between 50 Mpc and 16 Mpc, the later
corresponding to the small-scale gaussian filter
$r_s=5h^{-1}$ Mpc. Our result for the CDM model ${\cal M}
=1.2$ is in good agreement with the nonlinear calculation of
Suto, Cen, \& Ostriker (1992). The C+HDM model gives ${\cal
M} =2.2$, which perfectly agrees with the data quoted. If
instead of $L=58$ Mpc we choose $L=50$ Mpc, the results
would be slightly higher: ${\cal M} =1.4$ for CDM and ${\cal
M} =2.6$ for C+HDM.

Although the data is uncertain, as we said, C+HDM probably
does much better than CDM on the Mach number test.
Obviously, this test needs to be refined, and a number of
additional tests are needed, for example:

\item{$\bullet$} Galaxy velocities on small scales:
Analyses of velocities in the local group and other groups,
the cosmic virial theorem, and infall arguments typically
lead to $\Omega\approx 0.2-0.4$.  Are these results
consistent in detail with the predictions of $\Omega=1$
C+HDM?

\item{$\bullet$} Power spectra and velocities in spheres
compared with data:
To improve significantly on the linear predictions will
require large-scale simulations with high enough resolution
to identify galaxies.

\item{$\bullet$} Galaxy, cluster, and quasar formation and number
densities, ionization of IGM:
This could be addressed by high-resolution PM simulations
that stop at a redshift of order 1-2, but eventually will
require P$^3$M or tree codes including hydrodynamics and star
formation.  These calculations should also clarify the C+HDM
predictions for the evolution of the number densities of
galaxies, groups, and clusters as functions of their masses
or characteristic velocities, both in high- and low-density
regions.  New X-ray data on clusters (e.g. Edge \etal 1990,
Henry \& Arnaud 1991) could allow a critical test of the
model (see e.g. Kaiser 1991).

\item{$\bullet$} CMB anisotropies:
Current small-angle experiments should be close to detecting
temperature anisotropies if the C+HDM model is correct. The
compatibility of C+HDM with published microwave background
anisotropy results is discussed by Wright et al. (1992).
C+HDM predicts anisotropies for the OVRO experiment
(Readhead \etal 1989) and the MAX experiment (Devlin \etal
1992) that are near the upper limits derived from these
experiments. The model is incompatible with the result of
Gaier \etal (1992), who measure an upper limit to C(0) of
1.4e-5 from their South Pole experiment.  We note, however,
that this result was derived assuming a Gaussian form for
the radiation correlation function, which is not a
particularly good approximation to the function derived for
our model; further calculations are required to determine
whether this makes a significant difference in the derived
result. Moreover, this upper limit was obtained considering
only the channel with the least variation; a more
conservative treatment leads to limits that are compatible
with C+HDM (Dodelson \& Jubas 1993; J. R. Bond, private
communication). We also note that some fine tuning of the
model predictions for C+HDM can be done by changing the
amount of baryons in the model; for example, at a scale of
1.2 degrees the predictions for $\Delta T/T$ are reduced by
about 10\% if $\Omega_b$ is reduced from 0.10 to 0.05.

All other potentially viable cosmological models should
of course also be subjected to detailed confrontation with
data. Similar tests should therefore be applied to all the
main competetors to C+HDM, such as $\Omega=0.2$ CDM,
$\Omega=1$ CDM with designer spectra, or $\Omega=1$ cosmic
strings or other non-Gaussian fluctuation models with hot
dark matter.  Based on the results that have been obtained
thus far for C+HDM, we have reason to expect that it will do
very well.

\vfill\eject
\centerline{\bf Acknowledgments}

We thank Marc Davis, Avishai Dekel, Sandra Faber, Lars
Hernquist, Adrian Melott, Richard Schaeffer, and Sergei
Shandarin for valuable discussions. AAK wishes to
acknowledge support from NSF grant AST-902144 and NASA grant
NAGW-2923 at the University of Kansas. JRP and ER
acknowledge support from NSF grant PHY-9024920 at UCSC. Our
computer simulations were done on a grant of Cray-2 and
Convex-3 time at the National Center for Supercomputing
Applications.

\vfill\eject

\centerline{\bf References}

\ref    Achilli, S., Occhionero, F., \& Scaramella, R.; 1985; \apj; 299; 577.
\ref    Adams, F.C., Bond, J.R., Freese, K., Frieman, J.A.,
        \& Olinto, A.; 1992; \prd; 47; 426.
\ref    Babul, A., \& White, S.D.M.; 1991; \mnras; 251; P31.
\ref    Bacall, N.A., \& Cen, R.; 1992; \apjl; 398; L82.
\ref    Balian, R., \& Schaeffer, R.; 1988; \apjl; 335; L43.
\ref    Bardeen, J.M., Bond, J.R., Kaiser, N., \& Szalay, A.S.; 1986;
        \apj; 304; 15.
\ref    Bond, J.R., Cole, S., Efstathiou, G., \& Kaiser, N.;
        1992; \apj; 379; 440.
\ref    Bond, J.R., \& Szalay, A.; 1983; \apj; 276; 443.
\ref    Bouchet, F.R., Schaeffer, R., \& Davis, M.; 1991; \apj; 383; 19.
\ref    Carlberg, R.G., Couchman H.M.P., \& Thomas, P.A.; 1990; \apj;
        352; L29.
\ref    Cen, R., Gnedin, N.Y., Kofman, L.A., \& Ostriker,
        J.P.; 1992; \apjl; 399; L11.
\ref    Couchman, H.M.P., \& Carlberg, R.G.; 1992; \apj; 389; 453.
\ref    Davis, M., Summers, F.J., \& Schlegel, M.; 1992;
        Nature; 359; 393.
\ref    Davis, M., Peebles, P.J.E.; 1983; \apj; 267; 465.
\ref    Davis, M., Efstathiou, G.,  Frenk, C.S., White, S.D.M.; 1985;
        \apj; 292; 371.
\ref    Davis, M., Efstathiou, G.,  Frenk, C.S., White, S.D.M.; 1992;
        Nature; 356; 489.
\ref    Davis, R.L., Hodges, H.M., Smoot, G.F., Steinhardt,
        P.J., \& Turner, M.S.; 1992; \prl; 69; 1856.
\ps     Dekel, A. 1992, in {\it Observational Cosmology}, Milan 1992,
        ed. G. Chincarini \etal, in press
\ps     Dekel, A., Bertschinger, E., Yahil, A., Strauss, M.,
        \& Davis, M. 1992, Astrophys. J., submitted
\ps     Dekel, A., \& Rees, M.J. 1992, in prep.
\ps     Devlin, M. \etal 1992, in {\it Proc. NAS Colloquium
        on Physical Cosmology}, in press.
\ps     Dimopoulos, S., Hall, L.J., \& Raby, S. 1992, preprint LBL-32484.
\ref    Dodelson, S. \& Jubas, J.M.; 1993; \prd; 000; 000.
\ref    Edge, A.C., Stewart, G.C., Fabin, A.C., \& Arnaud,
        K.A.; 1990; \mnras; 245; 559.
\ref    Efstathiou, G., Davis, M., Frenk, C.S., \& White, S.D.M.; 1985;
        \apjs; 57; 241.
\ref    Efstathiou, G., \& Rees, M.J.; 1988; \mnras; 230; 5P.
\ps     Efstathiou, G., Sutherland, W.J., \& Maddox, S.J.
        1990, \mnras, 348, 705
\ref    Ellis, J., Lopez, J.L., \& Nanopoulos, D.V.; 1992; \plb; 292; 189.
\ref    Fang, L.Z., Li, S.X., \& Xiang, S.P.: 1984; \aa; 140; 77.
\ps     Frenk, C.S., White, S.D.M., Davis, M., \& Efstathiou, G. 1988;
        \apj, 327, 507.
\ref    Gaier, T. \etal; 1992; \apjl; 398; L1.
\ps     Gelb, J.M., 1992; Fermilab preprint
\ref    Gelb, J.M., Gradwohl, B., \& Frieman, J.A.; 1993; \apjl; 403; L5.
\ref    Groth, E.J., Juszkiewicz, R., \& Ostriker, J.P.; 1989; \apj; 346; 558.
\ref    Henry, J.P., \& Arnaud, K.A.; 1991; \apj; 372; 410.
\ps     Hockney, R.W., \& Eastwood, J.W. 1981, {\it Numerical
        simulations using particles} (New York: McGraw-Hill)
\ref    Holtzman, J.A.; 1989; \apjs; 71; 1.
\ref    Holtzman, J.A., \& Primack, J.R.; 1993; \apj; 000; 000.
\ref    Kaiser, N.; 1991; \apj; 383; 104.
\ref    Kaiser, N. \etal; 1991; \mnras; 252; 1.
\ref    Kates, R.E., Kotok, E.V., \& Klypin, A.A.; 1991; \aa; 243; 295.
\ref    Katz, N., Hernquist, L., \& Weinberg, D.H.; 1992; \apjl; 399; L109.
\ref    Klypin, A.A., \& Shandarin, S.F.; 1983; \mnras; 204; 891.
\ref    Klypin, A.A., \& Melott, A.L.; 1992; \apj; 399; 397.
\ps     Kofman, L. 1991, in  {\it Proc. IUPAP Conf. Primordial
        Nucleosynthesis and Evolution of Early Universe}, K. Sato, ed.
        (Dordrecht: Kluwer Academic Publishers), pp. 495
\ref    Krauss, L.M., \& White, M.; 1992; \prl; 69; 869.
\ref    Liddle, A.R., \& Lyth, D.H.; 1992a; \plb; 291; 391.
\ps     Liddle, A.R., \& Lyth, D.H. 1992b, preprint SUSSEX-AST 92/8-1
\ref    Lilje, P.; 1992; \apjl; 386; L33.
\ref    Lucchin, F., Mattarese, S., \& Mollerach, S.; 1992; \apjl; 401; L49.
\ref    Madau, P.; 1992; \apjl; 389; L1.
\ps     Nusser, A., \& Dekel, A. 1992, Astrophys. J., 391,
        443, and preprint submitted
\ref    Ostriker, J.P., \& Suto, Y.; 1990; \apj; 348; 378.
\ref    Park, C., Gott, J.R., \& da Costa, L.N.; 1992; \apjl; 392; L51.
\ps     Primack, J. 1991, in  {\it Proc. IUPAP Conf. Primordial
        Nucleosynthesis and Evolution of Early Universe}, K. Sato, ed.
        (Dordrecht: Kluwer Academic Publishers), p. 193
\ref    Readhaed, A.C.S. \etal; 1989; \apj; 346; 566.
\ref    Rowan-Robinson, M. \etal; 1990; \mnras; 247; 1.
\ref    Scaramella, R.; 1992; \apjl; 390; L57.
\ref    Schaefer, R.K., Shafi, Q., \& Stecker, F.W.; 1989; \apj; 347; 575.
\ref    Schaefer, R.K., \& Shafi, Q.; 1992; Nature; 359; 199.
\ref    Schaefer, R.K., \& Shafi, Q.; 1993; \prd; 47; 1333.
\ref    Shafi, Q., \& Stecker, F.W.; 1984; \prl; 53; 1292.
\ref    Smoot, G.F. \etal; 1992; \apjl; 396; L1.
\ref    Souradeep, T., \& Sahni, V.; 1992; Mod. Phys. Lett. A; 7; 3541.
\ref    Starobinsky, A.A.; 1979; JETP Lett.; 30; 719.
\ref    Strauss, M.A., Yahil, A., Davis, M., Huchra, J.P.,
        \& Fisher, K.; 1992; \apj; 397; 395.
\ref    Suto, Y., Cen, R., \& Ostriker, J.P.; 1992; \apj; 395; 1.
\ref    Taylor, A.N., \& Rowan-Robinson, M.; 1992; Nature; 359; 396.
\ref    Valdarnini, R., \& Bonometto, S.A.; 1985; \aa; 146; 235,
\ref    van Dalen, T., \& Schaefer, R.K.; 1992; \apj; 398; 33.
\ref    Vogeley, M.S., Park, C., Geller, M.J., \& Huchra, J.P.; 1992;
        \apjl; 391; L5.
\ref    Ueda, H., Itoh, M., \& Suto, Y.; 1993; \apj; 000; 000.
\ref    Walker, T.P., \etal; 1991; \apj; 376; 51.
\ps     Weinberg, S. 1972, {\it Gravitation and Cosmology}
        (New York: Wiley)
\ref    Wright, E.L. \etal; 1992; \apjl; 396; L13.
\ref    Zel'dovich, Ya.B.; 1970; \aa; 5; 84.

\vfill\eject

\centerline{\bf Figure Captions}

\noindent {\bf Figure \the\plA --}
The spectra of fluctuations at different redshifts in the
``hot'' (top panel) and ``cold'' (bottom panel) components.
Curves from top to the bottom correspond to redshifts
$z=0,5,10,15,20,25$. The dashed curves present the
approximations (Equations \the\eqA) to the spectra.

\noindent {\bf Figure \the\plB --}
Distribution of matter in a rectangular slice of 10 Mpc
thickness in the 200~Mpc box simulation. The slice was
chosen to pass through the highest density concentration in
the simulation. The density concentration is a rich galaxy
cluster in the top right corner of the slice.
(a) ``Cold'' particles (8\% of the total number).
(b) Points for which the density is higher than
$\rho_{\rm total} > 10\langle \rho _{\rm total}\rangle$
($1/6$ of the points).
(c) The distribution of 1015 ``galaxies'' in the slice.
``Galaxies'' were first preselected as maxima of density, and
then those with mass larger than $10^{11}\Msun$ are
presented. The area of a circle representing a ``galaxy'' is
proportional to its mass.

\noindent {\bf Figure \the\plM --}
Distribution of ``cold'' (left panel) and ``hot'' (right
panel) particles in a slice of $220\,{\rm kpc}\times
7\Mpc \times 7\Mpc$ in the 14~Mpc simulation. We show 85\%
of ``cold'' particles ($1.3\times 10^4$ particles) and
16.6\% of ``hot'' particles ($1.2\times 10^4$ particles).
Note that while all large condensations in the plot are seen
both in ``cold'' and ``hot'' particles (for example, the
large group in the left top corner, a few big ``galaxies''
in the filament, and the filament itself), small halos of
``cold'' particles do not find counterparts in ``hot''
particles (for example, two ``galaxies'' in the void at the
right middle part). The resolution for this simulation is
55~kpc, which is about 4 times smaller than the distance
between small tick marks.

\noindent {\bf Figure \the\plC --}
Evolution of the number density $N(>M)$ of dark halos with
mass larger than some limit (labeled in units of $\Msun$).
(In this and the next two figures, the vertical axis is
labeled in units of Mpc$^{-3}$, with $H_0=50$ km s$^{-}$
Mpc$^{-1}$.) The three bottom unmarked curves (mass limits
$3\times 10^{11}\Msun\!$ and higher) are for the simulation
with 50~Mpc box size. The marked curves correspond to the
number of objects in the 14~Mpc simulation. The dot-dashed
curves presents results for the Press-Schecter
approximation.

\noindent {\bf Figure \the\plD --}
Number density of halos per unit volume and per unit
logarithmic mass interval at the present epoch. The right
full curve presents the number of halos in the 50~Mpc box
simulation and the left full curve is for the 14~Mpc
simulation. The dashed curve presents the analytical
approximation Equation (\the\eqB). Results for the CDM model
are shown as the dot-dashed curve.

\noindent {\bf Figure 6 --}
Comparison of the number densities of ``galaxies'' in
the C+HDM and CDM models. The triangles represent
results of the CDM numerical simulations (box size 50~Mpc)
and the dashed curves are for the Press-Schecter
approximation with the parameter $\delta_c=1.60$. Full
curves are the Press-Schecter approximation for the C+HDM
model. Both models have the same biasing parameter $b=1.5$.
The dot-dashed curve represents the Press-Schecter
approximation for CDM with $b=2.5$, $\delta_c=1.60$, and
$M=10^{12}\Msun$.

\noindent {\bf Figure \the\plF --}
Evolution of the correlation function of the total
density in the 50~Mpc box simulation. The curves in the
plot are labeled by the expansion parameter $a$. At the
final moment the slope of the correlation function is about
$\gamma = 1.8$, but the correlation length $2.2 h^{-1}\Mpc$
is too small as compared to the correlation length of
galaxies. The dashed curve in this figure shows the
correlation function of the ``cold'' matter.  Results for
our CDM simulation are shown as the dot-dashed curve.

\noindent {\bf Figure \the\plE --}
Correlation function of ``cold'' particles at the final
moment for the 50~Mpc box (bottom dashed curve) and for the
200~Mpc box (bottom full curve). The top dashed and full
curves represent correlation functions of ``galaxies'' for
50~Mpc and 200~Mpc simulations correspondingly.
The dot-dashed line is the power law $\xi =
(5.5 h^{-1}/r)^{1.8}$.

\noindent {\bf Figure \the\plL --}
Influence of different effects on the correlation function.
(a) Simulation with 50~Mpc box. Full curves are for
thresholded density (10 for bottom and 25 for top curves).
The dashed and dot-dashed curves show results for
``galaxies'' found as maxima of the number of ``cold''
particles inside a sphere of 100~kpc radius. The dashed
curve is for ``galaxies'' more massive than $3\times
10^{10}\Msun$ (913 objects). The dot-dashed curve is for the
87 ``galaxies'' with mass larger than $3\times
10^{11}\Msun$.  The markers represent the correlation
function of galaxies in the CfA catalog (Davis and Peebles
1983).
(b) Effects of peculiar motions on the correlation function.
Bottom curves are for the 50~Mpc and top curves are for the
200~Mpc simulation. Correlation functions of 913 and 2831
``galaxies'' in redshift space (real space) are shown as
solid (dashed) curves. The markers represent the correlation
function in redshift space for galaxies in the CfA slices
(Vogeley \etal 1992).

\noindent {\bf Figure \the\plG --}
Components of the pair-wise velocities in the C+HDM
50\Mpc box simulation. The top panel shows $\langle
v_{21}\rangle$ for ``cold'' particles (dashed curve) and for
``galaxies'' (full curve). The dot-dashed curve shows where
the velocity is equal to the Hubble velocity: a pair on the
curve would have zero {\it proper} velocity. The bottom
panel presents $\sigma_\parallel=v_{\parallel, rms}$ (full
curves) and $\sigma_\perp=v_{\perp, rms}$ (dashed curves)
for ``cold'' particles (top two curves) and for ``galaxies''
(bottom two curves).

\noindent {\bf Figure \the\plI --}
Same as Figure~\the\plG, but for the CDM model.

\noindent {\bf Figure \the\plH --}
Velocity dispersion of ``galaxies'' (full curves) along the
line of sight as a function of projected distance in the
50~Mpc box simulations for the C+HDM (bottom panel) and CDM
(top panel) models. The dashed curves are for all ``cold''
particles and the dot-dashed curves represent the observed
behavior $340 (hr)^{0.13}\kms$. Only pairs of objects with
$\Delta V < 1000\kms$ are taken into account.

\noindent {\bf Figure 13 --}
Dependance of the projected pair-wise velocities on
different effects. The top panel is for the 200~Mpc box
simulation. The solid curve is the velocity dispersion
of ``galaxies'' along the line of sight for C+HDM, and the
dot-dashed curve shows the fit $\sigma =380
(hr)^{0.13}\kms$, which is compatible with the data.
The bottom panel shows results for
different selection criteria. The long-dashed curve shows
results for 481 ``galaxies'' with mass larger than
$10^{11}\Msun$. The short dashed curve presents results for
pairs with the relative distance smaller than $10 h^{-1}\Mpc$.
The dispersion for the cutoff in the velocity
difference $\Delta V < 1500\kms$ is shown as the dot-dashed
curve.

\noindent {\bf Figure \the\plK --}
Distribution of the total velocity of galaxies relative to
the rest frame in the 50~Mpc box simulation (913 objects) as
a function of orbital circular velocity. The curve presents
the averaged rms velocity. The error bars show formally
estimated statistical uncertainties.

\noindent {\bf Figure \the\plN --}
The density distribution function multiplied by density
for the C+HDM 50~Mpc box simulation. Values of $\rho P(\rho)
=\sum_i{\rho_i(\rho; \rho+\Delta\rho)}/\Delta\rho N,
N=256^3$ are shown. The full curves are labeled by the
expansion parameter $a$. The dashed line corresponds to the
power law $P(\rho)\propto \rho^{-3}$. The dot-dashed curve
is the fit Equation (\the\eqE).

\noindent {\bf Figure 16 --}
Bulk flow velocity vs. top-hat radius $R$ (for $H_0=50$)
from linear ($b=1.5$) calculations for C+HDM (solid curve)
and CDM (dashed), compared with POTENT data with Gaussian
smoothing radius 1200 km s$^{-1}$ from Dekel (1992).
The data points are not independant. The error bars on the
data points take into account errors of velocities of
galaxies and those introduced by POTENT. Typically the
errors are about 15\%. Because of statistical fluctuations
(the velocity of one sphere is measured) the theoretical
prediction looks like a wide band. We show the band for the
90\% confidence level (with this probability a mesurement
must be above the lower bound or below the upper bound). The
middle curves correspond to the mean expected values.

\bigskip
\noindent {\it Figures are available by mail from:
Prof. Joel Primack, Institute for Particle Physics, University of
California, Santa Cruz, CA 95064 U.S.A.
Email (internet): joel@lick.ucsc.edu;  Fax:(408) 459 3043.}

\bye